\documentclass[twocolumn,showpacs,preprintnumbers,amsmath,amssymb]{revtex4}
\usepackage{dcolumn}
\usepackage{bm}
\usepackage{graphicx}
\usepackage{amsmath}
\begin{document}

\title{Chaotic shock waves of a Bose-Einstein condensate}
\author{Wenhua Hai$^*$, \ \ Qianquan Zhu, \ \ Shiguang Rong}
\affiliation{Key Laboratory of Low-dimensional Quantum Structures and
Quantum Control of Ministry of Education, and
\\ Department of Physics, Hunan Normal University, Changsha 410081, China} 
\email{whhai2005@yahoo.com.cn}

\begin{abstract}

It is demonstrated that the well-known Smale-horseshoe chaos exists
in the time evolution of the one-dimensional (1D) Bose-Einstein
condensate (BEC) driven by the time-periodic harmonic or
inverted-harmonic potential. A formally exact solution of the
time-dependent Gross-Pitaevskii equation (GPE) is constructed, which
describes the matter shock waves with chaotic or periodic amplitudes
and phases. When the periodic driving is switched off and the number
of condensed atoms is conserved, we obtained the exact stationary
states and non-stationary states. The former contains the stable
`non-propagated' shock wave, and in the latter the shock wave
alternately collapses and grows for the harmonic trapping or
propagates with exponentially increased shock-front speed for the
antitrapping. It is revealed that existence of chaos play a role for
suppressing the blast of matter wave. The results suggest a method
for preparing the exponentially accelerated BEC shock waves or the
stable stationary states.

\end{abstract}


\maketitle

\textbf{}

A shock wave in a compressible classical fluid is characterized by a
steep jump in gas velocity and density due to the collisions of
particles. The physics and applications of the shock waves have been
thoroughly investigated during the last century in different fields
of physics \cite{Courant, Shugaev, Gurovich}. Very recently, a
different type of shock waves in a quantum fluid is explored
\cite{Salasnich, Hoefer, Shvartsburg, Brazhnyi, Kulikov, Damski},
where the quantum gas is a Bose-Einstein condensate (BEC) governed
by the Gross-Pitaevskii equation (GPE). Experiments and numerical
simulations \cite{Dutton,Simula,Perez} have depicted a BEC that
exhibits the traveling fronts with steep gradients of shock waves.
The corresponding analytical studies
\cite{Zak,Damski,Kamchatnov,Adbullaev} have also shown that a shock
wave could develop in the attractive or repulsive BECs. However, as
a nonlinear Schr\"{o}dinger equation the driven GPE allows existence
of the recognizably chaotic behavior \cite{Malomed,Reinhardt,
Thommen}. Therefore, various physical behaviors of the BECs may be
affected by the chaos, that leads to the chaotic BEC solitons
\cite{Martin,Elyutin, Chong1}, chaotic atomic populations
\cite{Adbullaev2, Lee, Hai}, chaotic quantum tunneling
\cite{Coullet, Xie}, chaotic Bogoliubov excitations \cite{Zhang},
and the chaotic BEC collapse \cite{Filho}. All the above theoretical
works are based on the analytical or numerical approximations to the
GPE, because of the nonintegrability of the system with external
potentials. The aim of this paper is to present an exactly
analytical evidence of the chaotic and oscillating shock waves in 1D
attractive or repulsive BECs driven by the time-periodic harmonic or
inverted-harmonic potentials, containing the most studied
harmonically trapped BEC with zero driving.

It is well-known that classical Smale-horseshoe chaos can exist in a
parametrically driven Duffing system with the cubic nonlinear and
harmonic or inverted-harmonic force \cite{ Parthasarathy,
Venkatesan}. The harmonic potential is a most widely used trapping
potential in the investigations of BECs \cite{Dalfovo, Leggett}. The
expulsive parabolic potential has also been applied to study the
time evolution of quantum tunneling \cite{Matsumoto}, deterministic
chaos \cite{Zurek, Haydock} and accelerated bright solitons
\cite{Khaykovich, Liang} (e.g. see Fig. 3B of Ref. \cite{Khaykovich}
for the attractive interaction) in BECs. The harmonic and
inverted-harmonic potentials can be time-dependent \cite{Xue,Castin,
Adbullaev3}. The presences of such potentials will break the
integrability of GPE and bring the difficulties to the exact
researches of BECs. Although for different initial and boundary
conditions the nonlinear GPE have many solutions, the exact solution
of GPE with harmonic or inverted-harmonic potential has not been
reported yet even for the non-driving cases. Particularly, we know
that a turbulent flow is a fluid regime characterized by chaotic,
stochastic property changes, and as an important physical behavior
the unstable and irregular quantum turbulence has been studied
\cite{Horng}. Therefore, the chaotic shock wave as a different type
of turbulent flow warrants further investigation.

In the present paper, we treat the time evolution of a quasi-1D BEC
created initially in a range near the potential center and driven by
the time-periodic harmonic or inverted-harmonic potential, and seek
the formally exact solution of the time-dependent GPE. By using the
exact solution we reveal that the well-known Smale-horseshoe chaos
exists in the matter shock waves with unpredictable amplitudes,
phases and wave fronts. When we switch off the periodic driving and
keep the number of condensed atoms, the exact solution becomes
explicit functions of spatiotemporal coordinates which contain the
stationary and non-stationary states. The stationary states include
the exact stable `non-propagated' shock wave. The non-stationary
states describe the exactly controllable shock waves which
alternately collapse and grow for the harmonic trapping and
repulsive interaction or propagate with exponentially increased
shock-front speed for the antitrapping and attractive interaction.
The suppression of chaos to the blast of matter wave is revealed.
The results can be observed experimentally \cite{Hoefer, Simula},
and supply a method for preparing the stable stationary state and
the exponentially accelerated shock waves which are similar to the
accelerated bright soliton \cite{Khaykovich}.

There are some different methods for the reductions from original 3D
GPE to quasi-1D GPE. The resulting equations contain the
nonpolynomial version \cite{L. Salasnich} and the cubic nonlinear
one \cite{Adbullaev3} which has been used in the regime of shock
wave \cite{Brazhnyi, Damski, Perez}. Assuming the transverse wave
function to be the ground state of a harmonic oscillator of
frequency $\omega_r$, the governing longitudinal GPE becomes the
cubic nonlinear equation $i\hbar \psi_t =- \frac{\hbar^2}{2m}\psi
_{xx} + [V'_{\alpha}(x,t)+ g'_{1d} |\psi|^2]\psi$, where $m$ is
atomic mass, $g'_{1d}=m \omega_r g_0/(2\pi \hbar)= 2 \hbar\omega_r
a_s$ is the quasi-1D atom-atom interaction intensity with $a_s$
being the $s$-wave scattering length, $V'_{\alpha}(x,t)=(\alpha
\frac 1 2m\omega_x^2+V'_1\cos \omega t)x^2$ denotes the harmonic
$(\alpha=1)$ and inverted-harmonic $(\alpha=-1)$ potentials of
frequency $\omega_x$ with the driving of strength $V'_1$ and
frequency $\omega$ \cite{Adbullaev3}. Taking an experimentally
suitable frequency $\omega_0$ as the units of frequencies $\omega_x,
\omega$ and normalizing the time, space and wave function with
$\omega_0^{-1},\ l_0=\sqrt{\hbar/(m\omega_0)}$ and $1/\sqrt{l_0}$,
the GPE becomes the dimensionless one
\begin{eqnarray}
&& i \psi_t =- \frac{1}{2}\psi _{xx} + [k_{\alpha}(t)x^2+ g_{1d}
|\psi|^2]\psi, \nonumber \\ && k_{\alpha}(t)=(\alpha \frac 1 2
\omega_x^2+V_1\cos \omega t).\ \ \
\end{eqnarray}
Here the interaction is reduced to $g_{1d}= 2 \omega_r
a_s/(\omega_0l_0)$ and the potential strength $V_1$ is normalized by
$\hbar \omega_0$. Throughout the paper we take $|g_{1d}|=0.4$ which
means $\hbar\omega_0=25m\omega_r^2a_s^2$ with $m,\ \omega_r$ and
$a_s$ determined by experiment.

Noticing that Eq. (1) can describe a symmetry BEC system in the
transformation $x\to -x$, we consider a real physical process in
which a BEC is created experimentally with the symmetry profile
between positions $\pm x_0=\pm L$ at initial time $t_0$, then the
condensed atoms propagate along $\pm x$ directions such that the
boundary coordinates $\pm x_0(t)$ are time-dependent. Letting the
number of condensed atoms conserve as $N$, we face the definite
problem with the initial data \cite{Hoefer}
\begin{eqnarray}
\psi(x,t_0)=\left\{\begin{array}{ll} \psi(x,t_0)\ \ \ \ for \ \ \
|x|\le L,
\\
0\ \ \ \ \ \ \ \ \ \ \  for\ \ \ \ |x|>L
\end{array}\right.
\end{eqnarray}
and the boundary-dependent normalization condition
$\int_{-x_0(t)}^{x_0(t)}|\psi(x,t)|^2dx=N$ for any $t$. For a
practical BEC the boundary density $|\psi(\pm L,t_0)|^2$ could be
made nonzero by using Feshbach resonances \cite{Perez} that leads to
the steep jumps in atomic density and flow velocity, which
characterize the feature of shock wave \cite{Hoefer} with the shock
front $\pm x_0(t)$ at which the front gradients $\psi_x(\pm x_0,t)$
is discontinuous. Although the initial data of $\psi(\pm L,t_0)$ and
$\psi_x(\pm L,t_0)$ cannot be measured accurately, in mathematics,
they are corresponded with an unique solution of Eq. (1) strictly.
Hence, if we construct an exact solution $\psi(x,t)$ which obeys Eq.
(2) and the normalization condition experimentally, it will describe
the exact shock wave of the system. For other initial and boundary
conditions, of course, one could obtain different type of solutions.
In fact, the accelerated bright soliton of the system has been
reported for the initial soliton state \cite{Khaykovich}.

We now give the formally exact solution of Eq. (1) as
\begin{eqnarray}
\psi(x,t)=[A(t)x+iB(t)]e^{i[a(t)+b(t)x^2]},
\end{eqnarray}
where $A(t),\ B(t),\ a(t),\ b(t)$ are real functions of time and
obey the coupled equations
\begin{eqnarray}
&&  \dot a+g_{1d}B^2=0,\ \ \  \dot b+2b^2=-g_{1d}A^2- k_{\alpha}(t),
\nonumber \\
&& \dot A+3Ab=0,\ \ \dot B +Bb=0.
\end{eqnarray}
The solution (3) can be easily proved by inserting Eqs. (3) and (4)
into Eq. (1) directly. The atomic density and flow velocity are
associated with the macroscopic wave function
$\psi(x,t)=\sqrt{\rho}\ e^{i\theta}$ through
\begin{eqnarray}
\rho(x, t)&=&|\psi(x, t)|^2=A^2x^2+B^2, \nonumber \\
v(x,
t)&=&\frac{\hbar}{m}\theta_x=\frac{\hbar}{m}\Big(2bx-\frac{AB}{\rho}\Big).
\end{eqnarray}
In order to evidence the solution (3), we have to solve Eq. (4) as
follows. From the third equation of Eq. (4) we get $b=-\dot A/(3A),\
\dot b=\dot A^2/(3A^2)-\ddot A/(3A)$. Applying these formulas to the
second one of Eq. (4) produces the decoupled equation
\begin{eqnarray}
\ddot A=3k_{\alpha}(t)A+3g_{1d}A^3+\frac{5\dot A^2}{3A}.
\end{eqnarray}
This is just the parametrically driven Duffing equation with a
`quadratic damping' term. It is well known that according to the
Melnikov chaos criterion for a certain parameter region the
Smale-horseshoe chaos exists in such a system \cite{ Parthasarathy,
Venkatesan}. To confirm the existence of chaos numerically, we adopt
the MATHEMATICA code
\begin{eqnarray}
&&T=2\pi/\omega;e[\{Anew_{-},vnew_{-}\}]:=\{A[T],v[T]\}/.Flatten
\nonumber
\\ &&[NDSolve[\{A'[t]==v[t],v'[t] ==5/3*v[t]^2/A[t] \nonumber \\
&& +3(0.5 \alpha \omega_x^2+V_1 Cos[\omega t])A[t]+3g_{1d}
A[t]^3,A[0]==Anew,\nonumber
\\
&& v[0]==vnew\},\{A,v\},\{t,0,T\}]];Do[pic_i=ListPlot\nonumber
\\ && [Drop[Nestlist[e,\{Random[Real,\{0.2,1.5\}],\nonumber \\
&&Random[Real,\{-0.8,0.8\}]\},6010],10],\{i,1,20\}]
\nonumber
\end{eqnarray}
with the parameters $g_{1d}=- 0.4, \ 3V_1=20.2$ and random initial
conditions $\{A(0)\in[0.2,\ 1.5],\ \dot A(0)\in[-0.8,\ 0.8]\}$ to
plot two groups of Poincar\'{e} sections on the `phase space' $(A,\
\dot A)$ respectively for (a) $\alpha \omega_x^2=0.8,\ \omega=6.0$
and (b) $\alpha \omega_x^2=-0.8,\ \omega=5.4$. We find some periodic
and chaotic orbits from each group consisting of $20$ Poincar\'{e}
sections. Here any Poincar\'{e} section is a discrete set of the
phase space points at every period of the external potential. In
Fig. 1a and Fig. 1b, we show the superpositions of $20$ Poincar\'{e}
sections for each group respectively, where Fig. 1a with $\alpha=1$
means the harmonic potential case and Fig. 1b with $\alpha=-1$ the
case of inverted-harmonic potential. In the both cases, if we change
$g_{1d}<0$ to $g_{1d}>0$, the orbits will tend to infinities for
some parameter sets or tend to zero for some other parameter sets.
The chaotic region of parameter space has not been found for the
repulsive interaction case yet.
\begin{figure}[htp] \center
\includegraphics[width=3.5in]{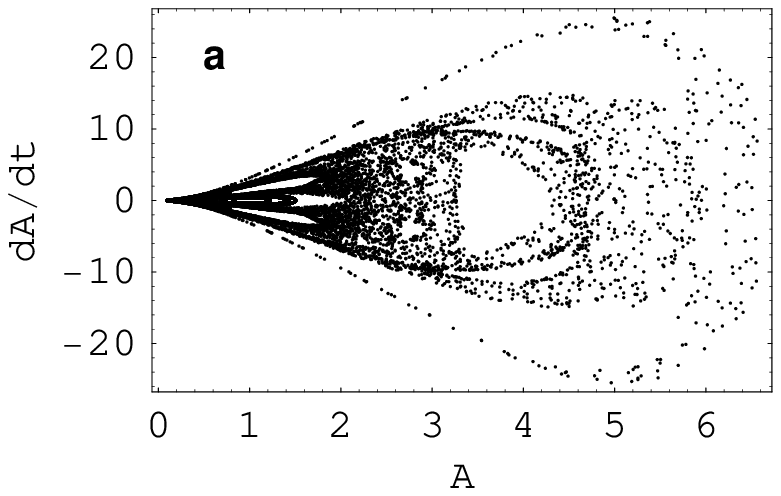}
\includegraphics[width=3.5in]{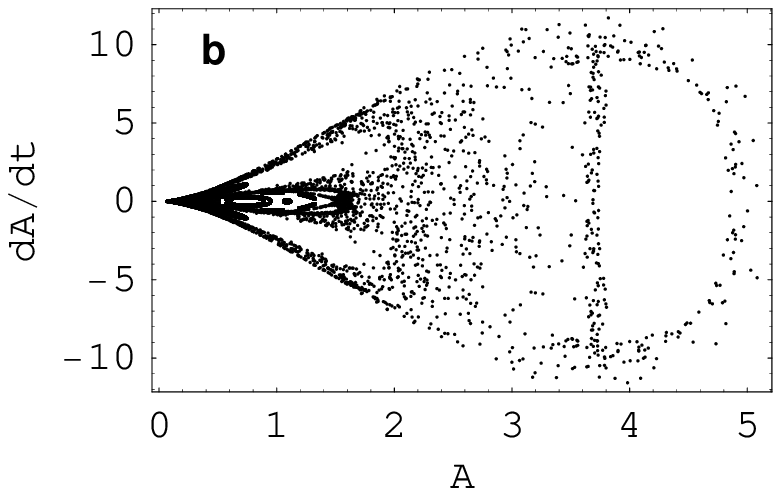}
\caption{The Poincar\'{e} sections on the dimensionless `phase
space' $(A,\ \dot A)$ for the parameters $g_{1d}=-0.4, \ 3V_1=20.2$
and (a) $\alpha \omega_x^2=0.8,\ \omega=6.0$, (b) $\alpha
\omega_x^2=-0.8,\ \omega=5.4$. Each figure consists of $20$
Poincar\'{e} sections associated with different initial conditions.}
\end{figure}
Given solution $A(t)$ from Eq. (6), the functions $B(t),\ a(t),\
b(t)$ can be derived from Eq. (4) immediately. Inserting them into
the exact solution of Eq. (3), the atomic density and phase gradient
of Eq. (5) become explicit forms, which govern the spatiotemporal
evolutions of the condensed atoms. The evolutions can be periodic
for $A(t)$ on the periodic orbits of Fig. 1 and can be chaotic for
$A(t)$ on the chaotic orbits of Fig. 1.

For the considered BEC created initially near the potential center,
the condensed atoms will propagate as a shock wave with
time-dependent wave front coordinate $x_0(t)$. At any fixed time the
atomic density profile of Eq. (5) is a parabola with the steep
gradients of shock fronts. The corresponding normalization condition
reads
\begin{eqnarray}
N=\int_{-x_0}^{x_0}|\psi(x,t)|^2dx=\frac 2 3 A^2(t)x^3_0+2B^2(t)x_0.
\end{eqnarray}
From the third and fourth equations of Eq. (4) we know the relation
$A=A_0B^3$ with constant $A_0$ such that Eq. (7) infers the formula
$2A_0^2[B^2(t)x_0]^3+6[B^2(t)x_0]=3N$, namely
$B^2(t)x_0=[A(t)/A_0]^{2/3}x_0=\lambda$ is a constant and the
coordinate $x_0(t)$ reads
\begin{eqnarray}
x_0(t)= \lambda [A_0/A(t)]^{2/3},
\end{eqnarray}
where $\lambda$ is fixed by the algebraic equation
$2A_0^2\lambda^3+6\lambda=3N$ and $A_0$ is determined by the initial
condition $x_0(t_0)=L$ in Eq. (2). It is interesting noting that
function $A(t)$ in Eq. (8) may be periodic or chaotic so that the
shock wave front $x_0(t)$ may periodically or chaotically oscillate.
For the periodic solution $A(t)$, the driven BEC breathes in the
spatiotemporal evolutions. In the chaotic solution case, the shock
wave front $x_0(t)$, BEC density $\rho(x,t)$ and flow velocity
$v(x,t)$ become unpredictable.

In the non-driving case with $V_1=0$ and $k_{\alpha}=\alpha \frac 1
2 \omega_x^2$, Eq. (6) becomes integrable. Applying the test formula
$\dot A^2=C_1A^2+C_2A^4$ to Eq. (6) and integrating this equation,
we obtain the undetermined constants $C_1=-\frac 9 2 k_\alpha,\
C_2=9g_{1d}$ and the two exact solutions
\begin{eqnarray}
A_{\alpha}(t)= \sqrt{\frac{k_\alpha}{2g_{1d}}}\ \textrm{sech}\Big[3
\sqrt{-\frac 1 2 k_\alpha}(t-t_0)\Big]
\end{eqnarray}
respectively for $g_{1d}<0,\ \alpha=-1$ and $g_{1d}>0,\ \alpha=1$,
where $t_0$ is an integration constant. The exact solutions of Eq.
(9) can be directly proved by substituting them into Eq. (6) with
constant $k_{\alpha}$. The $A_{-1}(t)$ of Eq. (9) denotes a
sech-shaped solution for the attractive interaction and
inverted-harmonic potential case. It is just the homoclinic solution
which is associated with the Smale-horseshoe chaos of Melnikov
criterion \cite{ Parthasarathy, Venkatesan} when the periodic
perturbation appears. This agrees with the numerical result shown in
Fig. 1. Noticing $\textrm{sech}(it)=\sec t$, the $A_{1}(t)$ means a
sec-shaped one for the repulsive interaction and harmonic potential
case. Given Eq. (9), from Eq. (4) we easily write the undetermined
functions
\begin{eqnarray}
B_{\alpha}(t)&=&\Big(\frac{A_{\alpha}}{A_0}\Big)^{1/3},\ \
b_{\alpha}(t)=-\frac{\dot A_{\alpha}}{3A_{\alpha}},\nonumber \\
a_{\alpha}(t)&=&-g_{1d}\int
\Big(\frac{A_{\alpha}}{A_0}\Big)^{2/3}dt,\ \ \alpha=\pm 1
\end{eqnarray}
with $A_0$ being an integration constant adjusted by the initial
conditions. Inserting Eq. (9) into Eq. (10), a simple calculation
can give the explicit forms of the functions $B_{\alpha}(t),\
a_{\alpha}(t)$ and $b_{\alpha}(t)$. Combining these functions with
Eq. (3), we obtain the corresponding explicit forms of the exact
non-stationary solutions immediately.

Applying Eq. (9) to Eq. (8), we find that for the sech-shaped
solution $A_{-1}(t)$ the shock wave front $x_0(t)$ is proportional
to $\cosh^{2/3}[\frac 3 2 \omega_x(t-t_0)]$ which increases
exponentially fast. However, for the sec-shaped solution $A_1(t)$
the shock wave front $x_0(t)$ is proportional to $\cos^{2/3}[\frac 3
2 \omega_x(t-t_0)]$ which oscillates periodically. The zero points
of the periodic $x_0(t)$ are just the singular points of $A(t)$,
where the BEC density $\rho(x,t)=A_1^2(t)x^2+[A_1(t)/A_{10}]^{2/3}$
becomes infinity. By inserting Eq. (3) into Eq. (1) and applying Eq.
(4) to the resulting equation, we find that at the singular points
of $A(t)$ the original GPE (1) is fulfilled, since the infinite
terms are offsetted each other. For a finite time the infinite
density means the escape of solution (3) and describes the alternate
collapses and growths of the blast matter wave physically
\cite{Hoefer, Simula}. In the both cases, the traveling fronts have
the steep gradients of shock waves \cite{Dutton,Simula,Perez}, since
normalization condition (7) implies that for a finite time
$\psi[x=\pm x_0(t),t]=$ \emph{complex constants} and
$\psi[|x|>|x_0(t)|,t]=0$ such that the front gradients $\psi_x[x=\pm
x_0(t),t]$ seem to be steep. No chaos exists for the non-driving
case.

Comparison between the driving and non-driving cases reveals that
existence of chaos could play a role for suppressing the escape of
solution and the blast of matter wave. In fact, from Fig. 1 we can
see that in the driving case with chaos, the function $A(t)$ is
finite and does not vanish for most of the orbits. These imply the
corresponding shock front $x_0(t)$ and density $\rho(x,t)$ having no
singular points.

For the time-independent `spring constant' $k_{\alpha}$, we have
also found that Eq. (4) has the simple solutions
\begin{eqnarray}
a=-\mu t, \ b=0, \ A^2=-\alpha \frac {\omega_x^2}{2g_{1d}}, \
B^2=\frac{\mu}{g_{1d}}.
\end{eqnarray}
Substituting Eq. (11) into Eq. (3) yields the exact stationary state
wave functions
\begin{eqnarray}
\psi_{\alpha}(x,t)=\Big[\pm\sqrt{-\frac{\alpha
\omega_x^2}{2g_{1d}}}\ x\pm i \sqrt{\frac{\mu}{g_{1d}}}\Big]e^{-i\mu
t}.
\end{eqnarray}
Here the harmonic potential with $\alpha=1$ is associated with the
attractive interaction $g_{1d}<0$ and negative chemical potential
$\mu$, and the inverted-harmonic potential with $\alpha=-1$
corresponds to the repulsive interaction $g_{1d}>0$ and positive
chemical potential. The second spatial derivative of Eq. (12)
vanishes that implies the zero kinetic energy of BEC and the
`non-propagated' shock wave with the steep front gradients. The
constant front $x_0$ is determined by the algebraic equation
$-\alpha \omega_x^2 x_0^3+6\mu x_0=3g_{1d}N$ from the normalization.
Such a non-propagated shock wave can be realized for the
time-independent potential $V=\alpha \frac{1}{2}\omega_x^2 x^2$ in
the region $ |x|\le x_0$ and $V=\alpha \frac{1}{2}\omega_x^2 x_0^2$
outside \cite{R.Dum}. Under such a finite potential, the stability
of BEC can be determined by the known criterion. For $\alpha>0$ and
$g_{1d}>0$, the stability criterion reads \cite{whai}
$\mu=\mu_s=V(x_0)+g_{1d}|\psi(x_0)|^2$. This implies that for the
harmonic potential and repulsive interaction the stationary state of
Eq. (12) is stable, since the $\psi(x_0)$ meets the stability
criterion.

In conclusion, for an atomic BEC created initially in a range near
the potential center and driven by the time-periodic harmonic or
inverted-harmonic potential, we have demonstrated that the classical
Smale-horseshoe chaos certainly exists in the time evolutions of the
system. The formally exact solution of the time-dependent GPE, whose
amplitude and phase depend on the solutions of the famous Duffing
equation with periodic driving and quadratic damping, has been
constructed, which describes the matter shock waves with chaotic or
periodic amplitude and phase. When the periodic driving is switched
off and the number of condensed atoms is conserved, we arrive at the
most studied BEC systems with parabolic potentials. Then the exact
solutions become the explicit functions of spatiotemporal
coordinates and govern the exact non-stationary states or the exact
stable stationary states. The stationary states are called the
`non-propagated' shock wave. The non-stationary shock waves
alternately collapse and grow for the harmonic trapping and
repulsive interaction or propagate with exponentially increased
shock-front speed for the antitrapping and attractive interaction.
It is revealed that existence of chaos play a role for suppressing
the blast of matter wave. The results can be observed experimentally
and suggest a useful method for preparing the exponentially
accelerated matter shock wave or the stable stationary state of the
BEC system.

The well-known criterion for the onset of temporal chaos is the
Melnikov criterion based on the nonlinear ordinary differential
equations. In order to apply such a criterion to demonstrate the
chaotic behaviors of Eq. (1), we have to consider the ansatz (3)
governed by the reduced ordinary differential equations (4) and (6).
In general, the investigation of spatiotemporal chaos also can be
performed directly from the nonlinear partial differential equation
(1), by using the Deissler-Kaneko criterion \cite{Deissler}, which
relies on the determination of the time evolution of a function
defined by the integral of the square modulus of the difference
between wave functions with nearby initial conditions \cite{Filho}.

\begin{acknowledgments}
This work was supported by the National Natural Science Foundation
of China under Grant Nos. 10575034 and 10875039.
\end{acknowledgments}

\end{document}